\documentclass[a4paper]{article}

\usepackage{t1enc}
\usepackage[latin1]{inputenc}
\usepackage[english]{babel}
\usepackage{amssymb}
\usepackage{latexsym}
\usepackage{amsmath,amssymb,mathdots}
\usepackage[enableskew]{youngtab}
\usepackage{graphicx}
\usepackage{epstopdf}
\usepackage{color}
\usepackage{authblk}

\usepackage{amsthm}
\usepackage{yfonts}

\usepackage{bbm}
\usepackage{bm}
\usepackage{mathrsfs}

\newcommand{\be}[0]{\begin{equation}}
\newcommand{\ee}[0]{\end{equation}}

\begin{document}
\title{\textbf{A new approach to the construction of Schur-Weyl states}}
\author[1]{Micha\l{} Kaczor \thanks{Email address: mkaczor.urinf@gmail.com}}
\author[1]{ Pawe\l{} Jakubczyk}
\affil[1]{University of Rzesz\'{o}w\\ College of Natural Sciences, Institute of Physics\\ Rejtana 16A, 35-959 Rzesz\'{o}w, Poland}
\date{\today}

\maketitle

\begin{abstract}
The Schur-Weyl states belong to a special class of states with a symmetry described by two Young and Weyl tableaux. 
Representation of physical systems in Hilbert space spanned on these states enables to extract quantum information hidden in nonlocal degrees of freedom.
Such property can be very useful in a broad range of problems in Quantum Computations, especially in quantum algorithms constructions, therefore it is very important to know exact form of these states.
Moreover, they allow to reduce significantly the size of eigenproblem, or in general, diminishing the representation matrix of any physical quantities, represented in the symmetric or unitary group algebra.
Here we present a new method of Schur-Weyl states construction in a spin chain system representation.
Our approach is based on the fundamental shift operators out of which one can build Clebsch-Gordan coefficients for the unitary group $U(n)$ and then derive appropriate Schur-Weyl state probability amplitudes.
\end{abstract}

\section{Introduction}
The structure of a quantum computer \cite{lul1} happens to have much in common with the Schur-Weyl duality \cite{lul2,lul3,lul4,TiagoCruz} between the unitary group $U(n)$ and the symmetric group $\Sigma_N$, acting on the $N$-th tensor power space $h^{\otimes N}$ of the defining space $h$ of $U(n)$. Explicitly, the space $h$ is identified with the elementary memory unit, referred to as a \textit{qunit} (a qubit for the case $n=2$), and then $h^{\otimes N}$ becomes the memory of a computer composed of $N$ such qunits, possibly subdivided between several parties, like Alice, the King, etc. \cite{lul5}. The memory space $h^{\otimes N}$, with dimension dim$\,h^{\otimes N}=n^N$, provides a variety of orthonormal bases, more or less adapted to several specific purposes of quantum information processing.
The calculational basis in the space $h^{\otimes N}$ is \textit{local}, i.e. each its element carries the exact information on the position of each particular qunit. It is, therefore, a fully separable state, whereas any transmission of information requires some entanglement. Processing with such entangled states is more efficient in some non-local bases of $h^{\otimes N}$, when it allows for an easy scan of information spread over a variety of qunits, in accordance with the quantum superposition principle. The best known, and perhaps the most radical way to display such non-local variables is \textit{the Fourier transform} over the set of qunits, which corresponds to change the position by momenta. These two sets of discrete variables, form the so called mutually unbiased bases \cite{lul6,lul7,lul8,lul9}, such that the full knowledge of quantum numbers specifying one basis exactly wipes out any information on the other. Bacon et al. \cite{lul10} have argued that the irreducible basis of the Schur-Weyl duality also provides a convenient access to non-local variables. They pointed out the importance of this basis in such prominent subjects of contemporary quantum information processing as universal quantum source coding \cite{lul10a}, communication without common reference frame \cite{lul11}, and any others, in which use of this basis is optimal. They also have proposed \cite{lul12} and demonstrated explicitly \cite{lul13} a method for determination of the Schur-Weyl basis in terms of a quantum circuit, obtained in a polynomial number of steps with respect to $N$ and $n$ ($n$ and $d$, respectively, in their notation).

The aim of the present paper is to propose the new method of construction of the Schur-Weyl states. The main advantage of our approach is the size of needed calculations, linear with respect to both $N$ and $n$. Moreover, we claim that our method is more transparent from the point of view of combinatorics, associated with the duality of Schur-Weyl and the relevant representation theory \cite{lul16,lul17}. 
We provide this transparency by a clear motivation of combinatoric entities at each step of the method. The main point which yields a simplification of the procedure consists in replacement of standard and semistandard Young tableaux (responsible for the irreducible bases of the symmetric and unitary group, respectively) by the double Gelfand patterns \cite{lul18,lul19}. 
These two sets carry the same combinatoric information, and the former is concise (and also closer to express the essence  of the Schur-Weyl duality at the level of bases), whereas the latter, being more extended, is flexible enough to indicate clearly the famous Robinson-Schensted-Knuth combinatoric algorithm \cite{robinson,schensted,knuth} as a path on the Gelfand pattern, resulting from the  ramification rules (betweenness conditions) in   a transparent geometric way. In particular, each Schensted insertion of a letter into an intermediate pattern is represented by a step (to the left or to the right) on this path.
To construct the Schur-Weyl states amplitudes we exploited the fundamental tensor operators, and 
combinatorial bijection between semistandard Weyl tableaux and Gelfand-Tsetlin triangles  \cite{lul18,gelfand,gelfand1}. We developed also a method of construction the directed graph   with the vertices labelled by Gelfand-Tsetlin triangles and the edges by single node states. This graph describes different scenarios of ladder construction of spin system "node by node", leading to the formula for the Schur-Weyl states amplitudes.

The paper is organised as follows. We start in Section 2 with a brief description of the $U(n)$ invariant physical model, the representation of Schur-Weyl duality and we introduce the
Schur-Weyl states.
In Section 3 we present the Robinson-Schensted-Knuth algorithm in the language of Gelfand-Tsetlin patterns. 
Section 4 is the main section where we present the algorithm of construction of the Schur-Weyl states probability amplitudes together with a simple example of calculation.
We end with  concluding remarks in Section 5.

\section{Schur-Weyl duality representation in one-dimensional spin system}

The memory space $h^{\otimes N}$ of a quantum computer is a scene of two groups linear actions: the symmetric group $\Sigma_N$, defined on the set
\be
\tilde{N}=\{j = 1,2,...,N\},
\ee
and the unitary group $U(n)$, defined on the qunit $h\cong\mathbb{C}^n$. The latter definition involves the set
\begin{equation}
\tilde{n}=\{i = 1,2,...,n\} 
\end{equation}
of labels of a unitary basis elements in $h$. $\Sigma_N$ and $U(n)$ are referred to \textit{dual groups}, and the corresponding \textit{dual sets}, $\tilde N$ and $\tilde n$, are usually referred to the \textit{alphabets}, of \textit{nodes} and \textit{spins}, respectively. These two alphabets define a basis
\begin{equation}\label{baza_konfiguracji}
\tilde n^{\tilde N} = \{ f: \tilde N \rightarrow \tilde n \}
\end{equation}
in the memory space $h^{\otimes N}$, such that each mapping  $f: \tilde N \rightarrow \tilde n$, referred to as a \textit{configurations of spins}, labels the pure state $|f\rangle \in h^{\otimes N}$ of the form
\begin{equation}\label{konfiguracja}
|f\rangle = | i_1 \rangle \otimes | i_2 \rangle \otimes \ldots \otimes | i_N \rangle \quad (i_j \in \tilde n,\mbox{ for }  j \in \tilde N).
\end{equation}
The set $ \tilde n^{\tilde N} $ is referred to as the \textit{initial}, or \textit{calculational} basis in $h^{\otimes N}$. Eq. (\ref{konfiguracja}) implies that any basis state $|f\rangle\in h^{\otimes N}$ is separable and, moreover, each qunit $h_j$, $j \in \tilde N$, is in a definite pure state $|i_j\rangle, \; i_j \in \tilde n$.

The Schur-Weyl duality \cite{lul2,schur_1927,weyl_1946} is presented in terms of actions, 
$A:\Sigma_N \times h^{\otimes N} \rightarrow h^{\otimes N}$ and $B:U(n) \times h^{\otimes N} \rightarrow h^{\otimes N}$, of two dual groups on the memory space $h^{\otimes N}$. We specify these actions in the calculational basis $ \tilde n^{\tilde N} $. The action $A$ of $\Sigma _N$ permutes the qunits along the formula
\begin{equation}\label{dzialanie_A}
A(\sigma) = {f \choose f \circ \sigma^{-1}}, \quad f \in \tilde n^{\tilde N}, \quad \sigma \in \Sigma_N,
\end{equation}
whereas the action $B$ of $U(n)$ transforms uniformly the entry of each qunit, which results in a multilinear transformation
\begin{equation}\label{dzialanie_B}
\begin{array}{l}
B(a)|f\rangle = \left(a|i_1\rangle\right)\otimes (a|i_2\rangle)\otimes \ldots \otimes (a|i_N\rangle) = 
\\
~~~~~~~~~~~~=\sum_{f'\in \tilde n^{\tilde N}} a_{i_1^{'}\, i_1} \ldots a_{i_N^{'}\, i_N} |f'\rangle, \quad f\in \tilde{n}^{\tilde{N}}, \quad a\in U(n)\\
\end{array}
\end{equation}
where
$a= (a_{i^{'}\, i} | i^{'}, i\in \tilde n) \in U(n)$, and $f'=(i_1^{'}, \ldots, i_N^{'}) \in  \tilde n^{\tilde N} $. These two actions mutually centralize, i.e.
\begin{equation}
A(\sigma) B(a) |f\rangle =  B(a) A(\sigma) |f\rangle, \quad \sigma \in \Sigma_N,\quad a\in U(n),\quad f\in  \tilde n^{\tilde N}, 
\end{equation}
what leads to a unique decomposition of the memory space
\begin{equation}
h^{\otimes N}= \bigoplus_{\lambda \in D(N,n)} \mathcal{H}^\lambda
\end{equation}
into sectors $\mathcal{H}^\lambda$, labelled by partitions $\lambda \in D(N,n)$. The set $D(N,n)$ denotes all partitions $\lambda$ of the integer $N$ into no more than $n$ parts, i.e. 
$\lambda=(\lambda_1, \ldots, \lambda_N), \; \lambda_1 \geq \lambda_2 \geq \ldots \geq \lambda_N \geq 0, \; \lambda_1+\lambda_2+ \ldots + \lambda_N = N$. Each sector $\mathcal{H}^\lambda$ is irreducible under the action of direct product group $U(n)\times \Sigma_N$, what leads to the decomposition
\begin{equation}\label{rozkladH}
\mathcal{H}^\lambda = V^\lambda \otimes W^\lambda,
\end{equation}
where $V^\lambda$ and $W^\lambda$ is the carrier space of the irrep $D^\lambda$ of $U(n)$ and $\Delta^\lambda$ of $\Sigma_N$, respectively. At the level of bases, it reads
\begin{equation}
D^\lambda(a) |V^\lambda\, t\rangle = \sum_{t' \in WT(\lambda,n)} D_{t' t}^\lambda (a) |V^\lambda\, t'\rangle, \quad t \in WT(\lambda,n),
\end{equation}
and
\begin{equation}
\Delta^\lambda(\sigma) |W^\lambda\, y\rangle = \sum_{y' \in SYT(\lambda)} \Delta_{y' y}^\lambda (\sigma) |W^\lambda\, y'\rangle, \quad y \in SYT(\lambda),
\end{equation}
for the irrep $D^\lambda$ and $\Delta^\lambda$, respectively, with the sets $WT(\lambda,n)$ and $SYT(\lambda)$ labelling the corresponding irreducible basis vectors. Thus 
$t \in WT(\lambda,n)$ is a Weyl tableau, or a semistandard Young tableau of the shape $\lambda$ in the alphabet $\tilde n$ of spins, and $y \in SYT(\lambda)$ is a standard Young tableau of the same shape $\lambda$ in the alphabet $\tilde N$ of nodes, whereas $D_{t' t}^\lambda (a)$ and $\Delta_{y' y}^\lambda (\sigma)$ denotes the relevant matrix elements. In this way, the duality (\ref{dzialanie_A})-(\ref{rozkladH}) imposes in the memory space $h^{\otimes N}$ the basis 
\begin{equation}\label{baza_SW}
\begin{array}{l}
b_{SW} = \left\{ |\lambda\, t\, y\rangle = |V^\lambda\, t\rangle \otimes |W^\lambda\, y\rangle \;\right\}
\\[5pt] \lambda \in D(N,n), \; t \in WT(\lambda,n), \; y \in SYT(\lambda),
\end{array}
\end{equation}
referred hereafter to as \textit{the Schur-Weyl basis}. 

It is worthwhile to compare these two bases, the initial $ \tilde n^{\tilde N} $ and that of Schur-Weyl $b_{SW}$, in the memory space $h^{\otimes N}$. To this purpose, one distinguishes two kinds of variables, internal and positional. The former are associated with the unitary degrees of freedom within a qunit $h$, whereas the latter relate to the labels of qunits within the memory space, and thus to the alphabet $\tilde N$ of nodes. The calculational basis is \textit{local}: for any configuration $f \in  \tilde n^{\tilde N} $, an individual qunit $j \in \tilde N$ is in a definite state $|i_j\rangle \in h$, $i_j \in \tilde n$. The Schur-Weyl basis is \textit{separable} with respect to these two kinds of variables: each Weyl tableau $t \in WT(\lambda,n), \; \lambda \in D(N,n)$, is associated with a \textit{collective} internal variable, composed as an $N$-th rank tensor along the distribution of letters of the alphabet $\tilde n$ of spins in the Weyl tableau $t$, whereas each standard Young tableau $y \in SYT(\lambda), \; \lambda \in D(N,n)$, involves a symmetrized combination of positional labels of qunits along the prescription coded in the standard Young tableau $y$. Clearly, the symmetrization procedure associated with the symmetric group $\Sigma_N$ implies that the Schur-Weyl basis is \textit{nonlocal}. It is worth to mention, however, that some information on localization of qunits still remains, in the form of ,,symmetrized localization'' of each letter $j \in \tilde N$ of the alphabet of nodes, seen as a box in the standard Young tableau $y$. This information on a symmetrized localization of qunits is fully reflected in the spectrum of Jucys-Murphy operators \cite{lul20,lul22,lul23,lul24} in the group algebra of $\Sigma_N$. A brief account of comparison between the two bases in the memory space $h^{\otimes N}$ is presented in Table \ref{tabela1}.
\begin{table}[h]
\caption{A comparision between calculational and Schur-Weyl bases in the memory space $h^{\otimes N}$.}
\centering
\begin{tabular}{l|l|l}
\hline
 ~~~~~~~~~~~$\setminus$ Basis & calculational & Schur-Weyl \\
Variable ~$\setminus$ & $ \tilde n^{\tilde N} =\{ f:\tilde N \rightarrow \tilde n \} $ & $ b_{SW} =\{ |\lambda\, t\, y \rangle \} $\\
\hline
internal & individual & collective (tensorial) \\
positional & localised & symmetrized \\
\hline
\end{tabular}
\label{tabela1}
\end{table}
 
 \subsection{The Schur-Weyl states}
 According to quantum mechanics, it is clear, that the elements of Schur-Weyl basis (\ref{baza_SW}) can be presented as linear combination of the calculational basis elements (\ref{baza_konfiguracji}) (cf. also Tab. \ref{tabela1})
\begin{equation}\label{stany_SW}
|\lambda \, t \, y \rangle \; = \;  \sum_{f \in \tilde n^{\tilde N}}
\langle f | \lambda t y \rangle \; |f \rangle.
\end{equation}
From this point of view, the Schur-Weyl basis elements $|\lambda\, t\, y\rangle$ can be seen as states  with amplitudes $\langle f | \lambda t y \rangle$ and symmetry described by the Weyl $t$ and Young $y$ tableaux.
To work with these states in optimal way, it is important to find their irreducible representation. To do this, let us consider  the action $A$ of the symmetric group in a purely combinatorial manner. This action decomposes the set of all magnetic configurations $\tilde n^{\tilde N}$ into orbits of the symmetric group
\begin{equation}\label{mh11}
\mathcal{O}_\mu = \{ f \circ \sigma^{-1} | \, \sigma \in  \Sigma_N \}, \;\;\; \mu \vDash N
\end{equation}
marked by $\mu$ \emph{- compositions} 
\footnote{A composition $\mu$ of a number $N$, $\mu \vDash N$ in short, is defined by a sequence of non-negative integers $(\mu_1, \mu_2, \ldots, \mu_n), \; \mu_i \in \mathbb{N}_{\geq 0}$ fulfilling the condition
$
 \sum_{i\in \tilde n} \mu_i = N.
$
}
of the number $N$.
Part $\mu_i$ of the composition $\mu$ are defined by
\begin{equation}\label{mh12}
\mu_i = |\{ i_j = i \, | \, j \in \tilde N  \}|, \,\,\, i \in \tilde n
\end{equation}
and corresponds to the number of nodes, occupied by appropriate state $|i\rangle, \; i \in \tilde n$ for each $f \in \mathcal{O}_\mu$.
Restriction of the action $A$ to the orbit $\mathcal{O_\mu}$ gives \emph{the transitive representation} of the group $\Sigma_N$
\begin{equation}\label{mh14}
\underbrace{A \big |_{\mathcal{O_\mu}}}_{\mbox{\scriptsize spanned on magetic configurations}} \equiv \underbrace{{R^{\Sigma_N:\Sigma^\mu}_{}}_{~~}}_{\mbox{\scriptsize spanned on left cosets of } \Sigma_N}
\end{equation}
with the stabiliser
$
\Sigma^\mu = \Sigma_{\mu_1} \times \Sigma_{\mu_2} \times \ldots \times \Sigma_{\mu_n}
$
being a Young's subgroup \cite{lul16,sagan}. This implies that $R^{\Sigma_N : \Sigma^\mu}$ can be spanned on the set of left cosets of the symmetric group $\Sigma_N$ with respect to the subgroup $\Sigma_\mu$, or on the set of  configurations of the orbit $\mathcal{O}_\mu$. Thus one can write
\begin{equation}\label{mh16}
|\mathcal{O}_\mu| = \frac{|\Sigma_N|}{|\Sigma^\mu|} = \frac{N!}{\prod_{i \in \tilde n} \mu_i !},
\end{equation}
i.e. the number of elements of the orbit $\mathcal{O}_\mu$ is equal to the number of cosets of the group $\Sigma_N$ with respect to the Young's subgroup $\Sigma^\mu$.

On the other hand, representation theory of the unitary groups uses the Kostka numbers $K_{\lambda \mu}$ ($\lambda,\; \mu$ are partitions) \cite{mcdonald, fulton} as the dimension of the carrier space of the representation $D^\lambda$ od $U(n)$, spanned on all Weyl tableaux of the weight $\mu$. Thus, the transitive representation can be decomposed into irreps of the symmetric group
\begin{equation}\label{roz_kostka}
R^{\Sigma_N : \Sigma^\mu} \cong \sum_{ \lambda \unrhd \mu} K_{\lambda \, \mu} \,\, \Delta^{\lambda},
\end{equation}
where the sum runs over the partitions $\lambda$ being greater than or equal to $\mu$ with respect to the order of domination
\footnote{
The partition $\lambda$ is greater than or equal to $\mu$ with respect to the order of domination, when
$
\lambda \unrhd \mu \Longleftrightarrow \sum_{i'=1}^{i} \lambda_{i'} \geq \sum_{i'=1}^{i} \mu_{i'}, \,\,\, i = 1, 2, \ldots ,
$
i.e. the sum of first $i$ parts of partition $\lambda$ is greater or equal than the respective sum of first $i$ parts of partition $\mu$ for every value of $i$.
},
$\Delta^{\lambda}$ stands for irrep of $\Sigma_N$ labelled by the partition $\lambda$.
The Kostka numbers $K_{\lambda \mu}$ in (\ref{roz_kostka}) are related to the (non-empty) intersection
\begin{equation}\label{k2}
lc_{\mathbb C} \mathcal O_\mu \cap \mathcal H^\lambda, \,\,\, \lambda \unrhd \mu
\end{equation}
of transitive representation space $R^{\Sigma_N : \Sigma^\mu}$ spanned on the orbit $\mathcal{O}_\mu$ with a sector $\mathcal H^\lambda$ of a space with the permutational symmetry $\lambda$. 
In another words, they represents the number of different copies of $V^{\lambda}$ inside the space spanned on the orbit $\mathcal{O}_\mu$, resulting in decomposition (\ref{rozkladH}).
The equation (\ref{roz_kostka}), written on the level of representations, can now be specified on the level of bases in a form
\begin{equation}\label{rozKostki}
|\mu \, \lambda \, t \, y \rangle \hspace{-3pt} = \hspace{-5pt} \sum_{f \in \mathcal{O}_\mu}
\langle \mu f | \lambda t y \rangle \;
|\mu f \rangle,
\end{equation}
where the probability amplitudes $\langle \mu f | \lambda t y \rangle$ allow for irreducible representation of  Schur-Weyl state $|\mu \, \lambda \, t \, y \rangle$ in terms of magnetic configurations.
We refer to Eq. (\ref{rozKostki}) as to the irreducible Schur-Weyl states, because it converts an initial base of  configurations $\mathcal O_\mu$ into the irreducible base
$
\{ |\mu \, \lambda \, t \, y \rangle \}
$
of the Schur-Weyl duality with the appropriate cross-section (\ref{k2}). 
The index $\mu$ in Eq. (\ref{rozKostki}) indicates that we restrict ourselves to the orbit $\mathcal O_\mu$ of the symmetric group.

The standard method of determination of the Schur-Weyl states amplitudes is based on the definition (see for example \cite{bohr_mottelson})
\begin{equation}\label{spr20}
|\lambda \, t \, y \rangle =\mbox{const} \sum_{\sigma \in \Sigma_N} \Delta_{y, y_t}^\lambda (\sigma) A(\sigma) |f_0\rangle,
\end{equation}
where $A(\sigma) |f_0\rangle = |f_0 \circ \sigma^{-1}\rangle$ and $\Delta_{y, y_t}^\lambda(\sigma)$ is appropriate matrix element of irrep $\Delta^\lambda$ for $\sigma \in \Sigma_N$. As we see in Eq. (\ref{spr20}), definition requires, roughly, $N$ factorial operations (because the sum runs over all elements of the symmetric group), thus practically, this method can be applied only to small systems consisting of at most a few dozen atoms. The method of Schur-Weyl states construction which we propose in following parts of the paper, is independent on the size of the system.

\section{Robinson-Schensted-Knuth algorithm for Gelfand-Tsetlin pattern}

The Robinson-Schensted-Knuth ({\rm RSK}) algorithm \cite{robinson,schensted} and its generalization \cite{knuth,knuth1998} has many applications, see for example its utilitarity in 
representation theory \cite{bjorner,kazhdan}, algebra \cite{mcdonald,sagan}, 
combinatorics \cite{fulton,stanley} 
and physics \cite{dorotaPawel2015,dorotaPawel2018}.
Originally, it establishes a bijective correspondence between the symmetric group elements and  pairs of Weyl and Young tableaux of equal shape.
In the spin system representation it provides a bijection
\begin{equation}
RSK : \tilde n^{\tilde N} \rightarrow b_{SW}
\end{equation}
between the initial basis $\tilde n^{\tilde N}$ (\ref{baza_konfiguracji}) of magnetic configurations and the irreducible basis $b_{SW}$ (\ref{baza_SW}) of the Schur-Weyl duality.

In this section we present a version of the RSK algorithm in the language of Gelfand-Tsetlin (GT) patterns  \cite{gelfand,gelfand1,louck2,louck3}. 
Generally speaking we substitute tableaux pairs by double Gelfand patterns \cite{lul18}.
These new irreducible basis elements carry the same combinatoric information, but are better adjusted to the Schur-Weyl duality approach and very well  reflect the physics of the spin systems \cite{lul18}.

\subsection{Gelfand-Tsetlin  patterns}
It is known that irreducible representations $D^\lambda$ of a unitary group $\mathrm{U}(n)$ are classified by a partitions $\lambda \in D_W(N,n) $. For consistency of further notation, such a partition is denoted as
\begin{equation}\label{gu1}
\lambda \equiv [m]_n=[m_{1n} \ldots m_{nn}].
\end{equation}
The standard basis of the carrier space $V^{[m]_n}$ of the irreducible representation $D^{[m]_n}$ is denoted by $\mathrm{GT}([m]_n, \tilde n)$, so that
\begin{equation}\label{gu11a}
V^{[m]_n}=lc_{\mathbb{C}} \; \mathrm{GT}([m]_n, \tilde n).
\end{equation}
Gelfand-Tsetlin patterns are adapted to the chain of unitary subgroups
\begin{equation}\label{gu2}
\mathrm{U}(1) \subset \mathrm{U}(2) \subset \cdots \subset \mathrm{U}(n-1) \subset \mathrm{U}(n),
\end{equation}
defined along of letters in the alphabet $\tilde n$ of spins. Consecutive restrictions of the irrep $D^{[m]_n}$ to subgroup $\mathrm{U}(i)$ along the chain (\ref{gu2}) (taken from the right to the left, i.e. $i=n,n-1, \ldots, 2,1$) are associated with partitions $[m_i]_i = (m_{1i} \ldots m_{ii})$, each corresponding to an irreducible representation $D^{[m_i]_i}$ of the intermediate subgroup $\mathrm{U}(i)$. Partitions $[m_i]_i$ can be arranged in an graphic way as

\begin{equation}\label{gu5}
\begin{array}{@{}llllllll|l@{}}
m_{1n} &          & m_{2n}   &           & \cdots  & m_{n-1 n} &             & m_{nn} & \mathrm{U}(n) \\

& m_{1n-1} &          & m_{2 n-1} &         & \cdots    & m_{n-1 n-1}      &&\mathrm{U}(n\!-\!1)     \\

&          & \ddots   &           &  \vdots &           & \iddots          && \vdots      \\

&          & m_{1 3}  &           & m_{2 3} &           & m_{33}          &&\mathrm{U}(3)      \\

&          &          & m_{1 2}   &         & m_{2 2}                       & &&\mathrm{U}(2)   \\

&          &          &           & m_{1 1}                                 && &&\mathrm{U}(1)   \\
\end{array}
\end{equation}
which is known as the Gelfand-Tsetlin pattern (or triangle). According to the Weyl ramification rule, each GT pattern $(m)$ with non-negative entries $m_{i, j}$ satisfies  \emph{the betweenness conditions}
\begin{equation}\label{gu1a}
m_{i-1 j} \leq m_{i-1, j-1} \leq m_{i,j}, \quad 1 \leq i \leq j \leq n.
\end{equation}
The first row of GT triangle coinciding with $[m]_n$, corresponds to a unique ray in $V^{[m]_n}$, so that the set of all GT patterns, $\mathrm{GT}([m]_n, \tilde n)$, yields an orthonormal basis for the irrep $D^{[m]_n}$ of $\mathrm{U}(n)$.
We choose the GT patterns $(m)$ instead of both semistandard Weyl tableaux $t$ or standard Young tableaux $y$, since the triangular shape (\ref{gu5}) of $(m)$ admits a transparent presentation of selection rules resulting from Weyl ramification. Namely, each row $i \in \tilde n$ of $(m)$ is an irrep of $U(i)$, which is indicated on the right side of the triangle (\ref{gu5}), and entries of the consecutive rows satisfy the betweenness condition (\ref{gu1a}).
Moreover, we write the basis state corresponding to $(m)$ in the form
\begin{equation}\label{gu7}
|(m)\rangle
=
\Bigg|
\left (
\begin{array}{@{}c@{}}
[m]_n \\
(m)_{n-1} \\
\end{array}
\right )
\Bigg \rangle,
\end{equation}
to make transparent the distinction between the label $[m]_n$ (square brackets) of the irrep of $U(n)$ (the first row of the triangle $(m)$), and its basis function $(m)_{n-1}$ (parentheses) - the remaining $(n-1)$ rows of $(m)$. In this notation, the
orthogonality condition for the basis $\mathrm{GT}([m]_n, \tilde n)$ reads
\begin{equation}\label{gu6}
\Bigg\langle {[m]_n \choose (m')_{n-1}} \Bigg| {[m]_n \choose (m)_{n-1}} \Bigg \rangle = \delta_{(m')_{n-1} (m)_{n-1}}
\end{equation}
and the dimension formula for $D^{[m]_n}$ as
\begin{equation}\label{gu10}
\mbox{dim} D^{[m]_n} = \frac{ \prod_{1 \leq i < j \leq n} (p_{i n}-p_{j n})}{1! 2! \ldots (n-1)!},
\end{equation}
where $p_{ij}= m_{ij}+j -i$ is known as the \emph{partial hook} corresponding to the $(i,j)$ entry of the GT pattern $(m)$.

We recall that the quantum state corresponding to the GT pattern $(m)$ can be presented, in a combinatorially equivalent way, by a semistandard Weyl tableau $t$ in the alphabet $\tilde n$ of spins as follows. Let the $i$ - th row of the tableau $t$ has the form 
$$
\underbrace{i \ldots i}_{\tau_{ii}}
\underbrace{i+1 \ldots i+1}_{\tau_{i,i+1}}
i+2 \ldots n-1
\underbrace{n \ldots n}_{\tau_{in}},
$$
so that $\tau_{ik}, 1 \leq i \leq k \leq n$, is the occupation number of the letter $k \in \tilde n$ in the $i$ - th row of $t$ ($\tau_{ik}=0$ for $i > k$ by advantage of semistandardness of $t$).
Then clearly
\begin{equation}\label{r27}
\sum_{i \in \tilde n} \tau_{ik} = \mu_k, \quad k \in \tilde n
\end{equation}
and
\begin{equation}\label{r28}
\sum_{k \in \tilde n} \tau_{ik} = \lambda_i, \quad i \in \tilde n
\end{equation}
determine the weight $\mu=(\mu_1, \ldots, \mu_n)$ and the shape $\lambda=(\lambda_1, \ldots, \lambda_n)$ of the tableau $t$. The equivalence between the tableau $t$ and the pattern $(m)$ is given by
\begin{equation}\label{r29}
\tau_{i,k} =
\left\{
\begin{array}{l}
m_{ik}-m_{i,k-1} \mbox{ for } 1 \leq i < k, \\
m_{ii} \mbox{ for } i = k,\\
0 \mbox{ for } i > k,\\
\end{array}
\right.
\end{equation}
together with the inverse transformation
\begin{equation}\label{r30}
m_{ik} = \sum_{1 \leq k' \leq k} \tau_{ik'}.
\end{equation}

\subsection{The Schensted insertion for Gelfand-Tsetlin patterns}

It is known that each semistandard Weyl tableau $t$ can be constructed recursively with respect to consecutive letters $j=1,2, \ldots, N$ of the alphabet $\tilde N$ of nodes, in accordance with the RSK algorithm. At the $j$ - th step of this recursion, $j \in \tilde N$, one applies the Schensted insertion, i.e. inserts a given letter $k \in \tilde n$ of the alphabet $\tilde n$ of spins into the intermediate tableau $t^{j-1}$ with the shape $\lambda^{(j-1)} = \mbox{shape}\,( t^{(j-1)})$ being a partition of $j-1$, along well known rules of the RSK algorithm. We adapt here these rules for application within the GT patterns $(m)$, equivalent to the tableau $t$ (cf. \ref{r30}). In order to insert the number $k$ into $(m)$ we follow the algorithm:
\begin{enumerate}
	\item
	mark the element $m_{1k}$ (the first element in the $k$ row of GT pattern) and increase it by one, i.e.
	$
	m_{1k} := m_{1k}+1;
	$

	\item
	then mark a subtriangle
	$$
	\begin{array}{ccc}
	m_{1 k+1} &   & m_{2 k+1}    \\
	& m_{1k} &     \\
	\end{array}
	$$
	where $m_{1k}$ is the element which has been increased in the first step;
	
	\item
	next check, if $m_{1k} > m_{1 k+1}$ then $m_{1 k+1} := m_{1 k+1} + 1$, in the opposite case $m_{2 k+1} := m_{2 k+1} + 1$;
	
	\item
	the element which has been increased becomes the starting point of a new subtriangle (like in step 2)
	$$
	\begin{array}{ccc}
	m_{i j+1} &   & m_{i+1 j+1}    \\
	& m_{ij} &     \\
	\end{array}
	$$
	where: $m_{ij}$ - element which has been increased by one in the previous step;
	
	\item
	and again, check
	if $m_{ij} > m_{i j+1}$ then $m_{i j+1} := m_{i j+1} + 1$ in opposite case $m_{i+1 j+1} := m_{i+1 j+1} + 1$;
	and, again, the element which has been increased becomes the new starting point of a new subtriangle;
	
	\item
	we repeat the procedure until we reach the $n$-th row of the GT pattern.
\end{enumerate}

The described procedure resembles some kinds of \emph{bubbling}, i.e. we obtain the travel path of the arguments $m_{ij}$ of the GT pattern, from row $k$ to $n$, along which the respective $m_{ij}$ are being increased by one.
\\~~\\

\noindent \textbf{Example}\\
Below we present the example of Schensted insertion for Gelfand-Tsetlin pattern. Suppose we have a triangle:
$$
{\footnotesize
\begin{array}{@{}lllllllll@{}}

7  &   & 4 &   & 2 &   & 1 &    & 0 \\

& 5 &   & 2 &   & 2 &   & 0      \\

&   & 5 &   & 2 &   & 0 &    &   \\

&   &   & 2 &   & 0 &   &    &   \\

&   &   &   & 2 &   &   &    &    \\
\end{array}
}
$$
and want to put the letter $k=2$ into it.
First we mark the element $m_{12}$, next we increase the value of this element by one and compare this element with $m_{13}$ and $m_{23}$.
Since $m_{12} \leq m_{13}$ $(3<5)$, we increase $m_{23}$ by one and compare it with $m_{24}$ and $m_{34}$.
Since $m_{23} > m_{24}$ (3 $>$ 2), we increase $m_{24}$ by one and compare it with $m_{25}$ and $m_{35}$, since $m_{24} \leq m_{25}$ $(3<4)$, we increase $m_{35}$ by one, and reach the top of the GT pattern. Finally, we obtain the new GT pattern
$$
{\footnotesize
\begin{array}{@{}lllllllll@{}}
7  &   & 4 &   & \boxed{3} &   & 1 &    & 0 \\
& 5 &   & \boxed{3} &   & 2 &   & 0      \\
&   & 5 &   & \boxed{3} &   & 0 &    &   \\
&   &   & \boxed{3} &   & 0 &   &    &   \\
&   &   &   & 2 &   &   &    &    \\
\end{array}
}
$$
where the rectangles mark the path of bubbling. The shape of the  bubbling path is strictly defined by the conditions of standardness of the triangle.
The resulting triangle represents the basis element of irrep $D^{(74210) + e_5(3)}$ of $U(5)$, where $e_5(3)=(0,0,1,0,0)$.

\subsection{Robinson-Schensted-Knuth algorithm}\label{rskalg}

In order to show explicitly that the above algorithm of insertion is bijective one has to construct the reverse  procedure. To achieve that, the number of row  gaining the new cell, at each step,  has to be coded. It can be resolved by adding an additional GT triangle which plays the role of the Young tableau in the classical RSK algorithm.
It leads to a double triangle of the form
\begin{equation}\label{rsg2}
\left (\begin{array}{c}
(y)_{n-1} \\
\mbox{[$m$]}_n\\
(t)_{n-1} \\
\end{array}
\right )
\end{equation}
which consists of two GT patterns with the same partition $\mbox{[$m$]}_n$ while the triangle $(y)_{n-1}$ is reflected in a horizontal plane. Symbol $(m)_{j}$ is used to mark rows from 1 to $j$ of the  GT pattern $(m)$, and $[m]_j$ reflects $j$-th row of the $(m)$. In other words, the lower triangle corresponds to the standard Weyl tableau and the top triangle - to the standard Young tableau from the classical RSK algorithm.
The RSK algorithm in terms of GT patterns for the spin configuration $f$ can be defined as follows:
\begin{enumerate}
\item[a)] write down the  configuration $f$ in a two-row notation
\begin{equation}\label{krs5}
f={1\; 2\; \ldots N\; \choose i_1 i_2 \ldots i_N}
\end{equation}
where the top row contains consecutive node numbers (alphabet of nodes) and the lower one - respective single-node states (alphabet of spins);
\item[b)]
draw the zero double GT pattern
\begin{equation}\label{rsg3}
\left (\begin{array}{c}
(0)_{n-1} \\
{[0]}_n\\
(0)_{n-1} \\
\end{array}
\right ),
\end{equation}
where symbol $[0]_n$ reflects row $n$ of a triangle consisting of $n$ zeros, i.e. $[\underbrace{0,0, \ldots, 0}_{n \mbox{{ \scriptsize times}}}]$;
\item[c)]
Then carry out the insertion of the successive elements (the insertion procedure described above) from the lower row of a configuration $f$ to a lower GT pattern; analogously form the top row of the configuration $f$ to the top GT pattern. 
However it must be outlined that, the insertion to the upper triangle starts with increasing an element $m_{i_1 i_2}$, where $i_1$ is equal to the number of partition parts $[m]_n$, which has been increased after the insertion of a lower letter, whereas  $i_2$ is a letter being inserted (from a top row of configuration (\ref{krs5})).
\end{enumerate}

\noindent
{\bf Example}
\\
Let us consider the spin configuration of the form
$$
|f\rangle = |31232\rangle = { 12345 \choose 31232}.
$$
We are looking for an appropriate double GT which is in one-to-one correspondence with $f$ according to the RSK algorithm. 
Firstly, we prepare an empty  configuration $|\emptyset \rangle$ and appropriate empty double GT pattern, next we insert consecutively letters from a configuration $|f\rangle$ in the following way:
\\\\
\footnotesize
$
\begin{array}{@{}lllllllll@{}}
&   &   &   & 0 &   &   &    &    \\
&   &   & 0 &   & 0 &   &    &   \\
&   & 0 &   & 0 &   & 0 &    &   \\
& 0 &   & 0 &   & 0 &   & 0      \\
0  &   & 0 &   & 0 &   & 0 &    & 0 \\
& 0 &   & 0 &   & 0 &   & 0      \\
&   & 0 &   & 0 &   & 0 &    &   \\
&   &   & 0 &   & 0 &   &    &   \\
&   &   &   & 0 &   &   &    &    \\
\end{array}
$
$
\begin{array}{c}
{1 \choose 3}\\
\rightarrow\\
\end{array}
$
$
\begin{array}{@{}lllllllll@{}}
&   &   &   & 1 &   &   &    &    \\
&   &   & 1 &   & 0 &   &    &   \\
&   & 1 &   & 0 &   & 0 &    &   \\
& 1 &   & 0 &   & 0 &   & 0      \\
1  &   & 0 &   & 0 &   & 0 &    & 0 \\
& 1 &   & 0 &   & 0 &   & 0      \\
&   & 1 &   & 0 &   & 0 &    &   \\
&   &   & 0 &   & 0 &   &    &   \\
&   &   &   & 0 &   &   &    &    \\
\end{array}
$
$
\begin{array}{c}
{2 \choose 1}\\
\rightarrow\\
\end{array}
$
\\[8pt]
$
\begin{array}{@{}lllllllll@{}}
&   &   &   & 1 &   &   &    &    \\
&   &   & 1 &   & 1 &   &    &   \\
&   & 1 &   & 1 &   & 0 &    &   \\
& 1 &   & 1 &   & 0 &   & 0      \\
1  &   & 1 &   & 0 &   & 0 &    & 0 \\
& 1 &   & 1 &   & 0 &   & 0      \\
&   & 1 &   & 1 &   & 0 &    &   \\
&   &   & 1 &   & 0 &   &    &   \\
&   &   &   & 1 &   &   &    &    \\
\end{array}
$
$
\begin{array}{c}
{3 \choose 2}\\
\rightarrow\\
\end{array}
$
$
\begin{array}{@{}lllllllll@{}}
&   &   &   & 1 &   &   &    &    \\
&   &   & 1 &   & 1 &   &    &   \\
&   & 2 &   & 1 &   & 0 &    &   \\
& 2 &   & 1 &   & 0 &   & 0      \\
2  &   & 1 &   & 0 &   & 0 &    & 0 \\
& 2 &   & 1 &   & 0 &   & 0      \\
&   & 2 &   & 1 &   & 0 &    &   \\
&   &   & 2 &   & 0 &   &    &   \\
&   &   &   & 1 &   &   &    &    \\
\end{array}
$
$
\begin{array}{c}
{4 \choose 3}\\
\rightarrow\\
\end{array}
$
\\[8pt]
$
\begin{array}{@{}lllllllll@{}}
&   &   &   & 1 &   &   &    &    \\
&   &   & 1 &   & 1 &   &    &   \\
&   & 2 &   & 1 &   & 0 &    &   \\
& 3 &   & 1 &   & 0 &   & 0      \\
3  &   & 1 &   & 0 &   & 0 &    & 0 \\
& 3 &   & 1 &   & 0 &   & 0      \\
&   & 3 &   & 1 &   & 0 &    &   \\
&   &   & 2 &   & 0 &   &    &   \\
&   &   &   & 1 &   &   &    &    \\
\end{array}
$
$
\begin{array}{c}
{5 \choose 2}\\
\rightarrow\\
\end{array}
$
$
\begin{array}{@{}lllllllll@{}}
&   &   &   & 1 &   &   &    &    \\
&   &   & 1 &   & 1 &   &    &   \\
&   & 2 &   & 1 &   & 0 &    &   \\
& 3 &   & 1 &   & 0 &   & 0      \\
3  &   & 2 &   & 0 &   & 0 &    & 0 \\
& 3 &   & 2 &   & 0 &   & 0      \\
&   & 3 &   & 2 &   & 0 &    &   \\
&   &   & 3 &   & 0 &   &    &   \\
&   &   &   & 1 &   &   &    &    \\
\end{array}
$
~~\\\\[12 pt]
And finally
\\
$$
{12345 \choose 31232} \longleftrightarrow
\begin{array}{@{}lllllllll@{}}
&   &   &   & 1 &   &   &    &    \\
&   &   & 1 &   & 1 &   &    &   \\
&   & 2 &   & 1 &   & 0 &    &   \\
& 3 &   & 1 &   & 0 &   & 0      \\
3  &   & 2 &   & 0 &   & 0 &    & 0 \\
& 3 &   & 2 &   & 0 &   & 0      \\
&   & 3 &   & 2 &   & 0 &    &   \\
&   &   & 3 &   & 0 &   &    &   \\
&   &   &   & 1 &   &   &    &    \\
\end{array}
$$
\normalsize
\\

According to (\ref{r29}) obtained double GT pattern is mapped to tableaux ${\scriptsize (\Yvcentermath1 \young(122,33), \young(134,25))}$.
One can easily check, that the classical RSK algorithm applied to $|f\rangle = |31232\rangle$ leads to the same result.

We have shown that the RSK algorithm can be fully realized in the GT basis which is better adapted to the symmetry of the spin systems.
This basis expresses all the rules of choice imposed by both dual groups $\Sigma_N$ and $\mathrm{U}(n)$, with the help of simple geometrical limitations of arguments of the GT pattern (compare the betweenness conditions (\ref{gu1a})). 
The double GT patterns are capable to substitute a pair $(t,y)$ of Weyl and Young tableaux completely. 
Proposed approach along with a ladder construction of spin nodes, are key elements in the construction of the Schur-Weyl states amplitudes.

\section{Construction of the Schur-Weyl states amplitudes}

\begin{figure}[h]
\begin{center}
\includegraphics[width=0.45\textwidth]{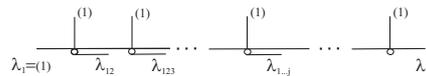}
\end{center}
\caption{Scheme of a ladder coupling of consecutive nodes of spin chain. $\lambda_{1..j}=\mbox{shape}\, t_{1..j}$ standing for the shape of tableau $t_{1..j}$ during the $j$-th step of coupling, whereas $(1)$ defines the cell of a tableau (single node state).} \label{skladanie}
\end{figure}

The amplitudes $\langle f | \lambda t y \rangle$ of the state (\ref{rozKostki}) can be calculated using the ladder construction (see Fig. \ref{skladanie})
of single nodes spin states $1,2, \ldots, j,j+1, \ldots N$ of the system. To maintain the symmetry, these couplings should be adjustments to the combinatorial growth of the Weyl tableau $t$ according to the RSK algorithm.
The addition of one node to the existing system prepared in a state $|\lambda_{1\ldots j-1}\, t_{1\ldots j-1} \rangle$ can be described in term of the Wigner-Clebsch-Gordan coefficient of the form
\begin{equation}\label{wspWCG}
{\scriptsize
\left [
\begin{array}{ccc}
\lambda_{1\ldots j-1} & (1) & \lambda_{1\ldots j}\\
t_{1\ldots j-1}  & f(j) & t_{1\ldots j}
\end{array}
\right ].
}
\end{equation}
This coefficient is responsible for the add of the $j$-th node (in the single-node state $f(j)$ - represented by the second column) to an existing system which consists of $j-1$ nodes (in states $t_{1\ldots j-1}$ represented by the first column), the resulting system consists of $j$ nodes (in states $t_{1\ldots j}$ - represented by the third column).
Addition of nodes can be considered as the one-by-one process, therefore one can replace the coefficient (\ref{wspWCG}) by matrix elements of a fundamental tensor operator \cite{lul18,louck1,louck4,louck4a} using the relation
\begin{equation}\label{fop}
\left [
\begin{array}{ccc}
\lambda_{1\ldots j-1} & (1) & \lambda_{1\ldots j}\\
t_{1\ldots j-1}  & f(j) & t_{1\ldots j}
\end{array}
\right ] = \langle t_{1..j} |  \hat F_{f(j), row(\lambda_{1..j} \setminus \lambda_{1..j-1})}  | t_{1..j-1} \rangle
\end{equation}
where $row(\lambda_{1..j} \setminus \lambda_{1..j-1})$ denotes the row number of the cell, which remain after deleting cells $\lambda_{1..j-1}$ from shape $\lambda_{1..j}$, $\hat F_{pq}$ is a fundamental tensor operator for the unitary group $U(n)$.

In general, construction of the amplitude $\langle f | \lambda t y \rangle$ is imposed by the decreasing process of the Weyl tableau $t$ during the reverse RSK algorithm applied to state $| \lambda t y \rangle$. 
Each single step of this process, from $\lambda_{1\ldots j}$ to $\lambda_{1\ldots j-1}$, obeys all selection rules involved in reversed RSK algorithm at every stage, and thus contributes additively to the total value of the amplitude.  
This process can be described in terms of graph $\Gamma$, which gives systematic description of all the possible ways of growth of the Weyl tableau $t$ out of magnetic configuration $f$.

Such a graph is simple and directed with minimal (initial) vertex equal to the zero Gelfand-Tsetlin pattern (i.e. a triangle of the shape lambda, filled with zeros), and maximal (final) vertex equal to the pattern which correspond (bijectively) to the Weyl tableau $t$.
Formally, graph $\Gamma$ consists of a set $GT$ of Gelfand-Tsetlin patterns as vertices and the set $\{f(i): i =1 \ldots N\}$ of single-node states which labels the edges (or arcs), such that $\Gamma = (GT, \{f(1), f(2), \ldots, f(N)\})$.
Edge $f(j)=(t_{12..j-1}, t_{12..j})$ of two adjacent vertices $(t_{12..j-1}, t_{12..j})$, where $t_{12..j-1}$ is the initial and $t_{12..j}$ is the terminal vertex of the edge $f(j)$, constructed by inserting the single node state (the letter ) $f(j)$  into the initial vertex $t_{12..j-1}$ in such a way that we obtain the state $t_{12..j}$.

The construction process of the graph $\Gamma$ can be split into two stages.
Firstly, as we mentioned above, we read off the sequence of the partitions $\lambda_{RS} = (\lambda = \lambda_{12\ldots N}=[m]_n, \lambda_{12\ldots N-1}=[m]_{N-1}, \ldots, \lambda_{12}=[m]_2, \lambda_{1}=[m]_1)$ from inverse RSK algorithm applied to the state $|\lambda t  y\rangle$, where $[m]_j$ is the $j$-th row of Gelfand-Tsetlin pattern (see Sect. (\ref{rskalg})), and $\lambda_{12\ldots j}$ is the shape of the Weyl (or Young) tableau at $j$-th step of reverse RSK algorithm.
Secondly, we construct the graph by \emph{insertion}, one by one, the consecutive letters $f(j)$ of configuration $f=f(1) f(2) \ldots f(N)$ to the Gelfand-Tsetlin patterns, starting from a triangle consisting of zeros only.
Insertion of one letter $f(j)$ into the Gelfand-Tsetlin pattern $t_{1..j-1}$ increase by one appropriate elements located in the rows $j$,  $f(j)~\leq~j~\leq~n$, i.e.
\begin{equation}\label{k8}
\left (
\begin{array}{c}
  [m]_n + e_n(\tau_n) \\
  \mbox{[$m$]}_{n-1} + e_{n-1}(\tau_{n-1}) \\
  \vdots \\
   \mbox{[$m$]}_{f(j)} + e_{f(j)}(\tau_{f(j)})\\
  (m)_{{f(j)}-1}\\
   \end{array}
 \right )
\end{equation}
where $\tau_j \in \{ 1,2, \ldots j \}$, and $[m]_j + e_j(\tau_j)$ denotes $j$-th row of the Gelfand-Tsetlin pattern, $e_j(\tau_j)$ is vector of zeros of the length $j$ with $1$ at the position $\tau_j$.
To calculate the probability amplitude of the adding the node $j$ prepared in state $f(j)$ to the system consisting of $j-1$ nodes prepared in the state $t_{1..j-1}$ we use a fundamental tensor operators. These operators, in the Gelfand-Tsetlin basis representation, can be calculated using a technique called \emph{pattern calculus}. This is used to determine matrix elements of tensor operators of any unitary groups with the help of symbolic diagrams and appropriate processing rules. Pattern calculus approach converting many complicated dependencies between the arguments of a vector state into \emph{obvious} geometrical limitations (betweenness conditions). Louck in \cite{lul18} has shown that this kind of fundamental tensor operator can be calculated using the formula
\begin{equation}\label{fso}
\begin{array}{l} 
\left.
 \begin{picture}(1,1)
  \raisebox{2pt}{ \put(0,0){\line(1,3){13}}  \put(0,0){\line(1,-3){13}} }
\end{picture}
\;\;\;
\begin{array}{c}
  \mbox{[$m$]}_n + e_n(\tau_n) \\
  \mbox{[$m$]}_{n-1} + e_{n-1}(\tau_{n-1}) \\
  \vdots \\
   \mbox{[$m$]}_k + e_k(\tau_k)\\
  (m)_{k-1}\\
   \end{array}
 \right |
 \hat F_{k, \, \tau_n}
\left |
\begin{array}{l@{}}
  [m]_n  \\
  \mbox{[$m$]}_{n-1} \\
 \vdots \\
   \mbox{[$m$]}_k \\
  (m)_{k-1}\\
   \end{array}
 \right.\;\;\;\;\,
 \begin{picture}(1,1)
  \raisebox{2pt}{ \put(0,0){\line(-1,3){13}}  \put(0,0){\line(-1,-3){13}} }
\end{picture}
 = 
\prod_{j=k+1}^{n} \mbox{sgn}(\tau_{j-1} - \tau_j)
\\ \\ 
\sqrt{
\left |
\frac
{\mathop{\prod_{i=1}^{j-1}}_{i \neq \tau_{j-1}}
\prod_{i=1, i \neq \tau_{j-1}}^{j-1} (p_{\tau_j,j} - p_{i,j-1})
 \prod_{i=1, i \neq \tau_{j}}^{j} (p_{\tau_{j-1},j-1} - p_{i,j}+1)}
{\prod_{i=1, i \neq \tau_{j}}^{j} (p_{\tau_j,j} - p_{i,j})
 \prod_{i=1, i \neq \tau_{j-1}}^{j-1} (p_{\tau_{j-1},j-1} - p_{i,j-1}+1)}
\right |
}
\cdot
\\
\sqrt{
\left |
\frac
{\prod_{i=1}^{k-1} (p_{\tau_k,k} - p_{i,k-1})}
{\prod_{i=1, i \neq \tau_{k}}^{k} (p_{\tau_k,k} - p_{i,k})}
\right |
}
\end{array} 
\end{equation}
for $k \in \{2,3, \ldots, n-1 \}$. 
If $k=n$ the first factor of the rhs of Eq. (\ref{fso})
is equal to 1; while for $k=1$ the second factor of the rhs of Eq. (\ref{fso})
is equal to 1.
Here $k=f(j)$ denotes the added node.
The partial hook $p_{ij} = m_{ij}+j-i$, $e_i(j)$ is the unit vector of the length $i$ with $1$ on the position $j$,  $[m]_i$ represents $i$-th row of Gelfand-Tsetlin pattern $(m)$, whereas $(m)_i$ denotes rows from 1 to $i$ of pattern $(m)$.
Equation (\ref{fso}) allows to express any matrix element of any fundamental tensor operator in a basis of Gelfand-Tsetlin patterns.

It is obvious that operation (\ref{k8}) can leads to a collection of patterns (because $\tau_j \in \{ 1,2, \ldots j \}$), but we choose only those, for which the $n$-th row $[m]_n + e_n(\tau_n)$ is equal to a partition $\lambda_{12..j}\in \lambda_{RS}$ (or in other words, the shape of a resulting Gelfand triangle should coincide with $\lambda_{12..j}$ the intermediate partition of the RSK algorithm), and the standardness of the Gelfand-Tsetlin pattern is conserved (or in other words, the betweenness conditions are satisfied).
In the language of graphs, this means that, the out-degree of the vertex (i.e. the number of edges coming from vertex) $t_{1..j-1}$, what we denote $deg^+(t_{1..j-1})$, can be greater than (or equal to) one.

Using the above insertion procedure, we start to insert the first letter $f(1)$ into the zero Gelfand triangle $t_0$ (i. e. a triangle of the shape $\lambda$, filled in by zeroes, which is the minimal vertex of our graph), what results in reaching the vertex $t_1$. This leads to a directed graph, consisting of two vertices $(t_0, t_1)$, joined by the edge $f(1)$
$$
\begin{array}{ccc}
&(t_0)&\\
&\downarrow &{\tiny f(1)}.\\
&(t_1)&\\
\end{array}
$$
Next, one inserts the letter $f(2)$ into the triangle $t_1$ leading to a set of vertices $t_{12}=\{t_{12}^i \; : \; i=1,2,...\}$. Geometrically this represents a graph exhibiting branches, with $deg^+(t_1) \geq 1$.
$$
\begin{array}{ccccc}
&&(t_0)&&\\
&&\;\;\;\;\;\;\;\; \downarrow {\tiny f(1)}&&\\
&&(t_1)&&\\
\;\;\;\;\;\;\;\;\;\;\;\; \swarrow f(2)&\ldots &\;\;\;\;\;\;\;\;\downarrow  f(2)& \ldots &\\
(t_{12}^1)&\ldots &(t_{12}^k)& \ldots & \ldots \\
\end{array}
$$
This is followed by further insertion of the letter $f(3)$ into each vertex $t_{12}^i$ from the set $t_{12}$ using the same rules, which produces the set $t_{123}$ of vertices composed of three letters. The same routine is followed for all remaining letters of the configuration $f$. 

One can observe that the out-degree $deg^+(t_{12..j})\geq 1$, and can be seen that the insertion rules themselves suggest a quick growth of the graph in a tree-like manner. Nevertheless, symmetry constraints imposed by the physical system, guarantee that final graph will result in the shape of a rhomb (see example below) with the maximal (final) vertex (resulting from the insertion of the last letter $f(N)$ of the configuration $f$) equal to $\lambda t$.

Our approach to calculation of amplitudes out of the graph $\Gamma$ resembles the $n$ slit interference experiment with electrons. In this experiment the calculation of probability amplitude for the transition of an electron, from a source $s$ through a sequence of walls, with slits in them, to the detector $x$, is given by the formula
\be\label{nslitExper}
\langle x|s\rangle = \sum_{\mbox{{\scriptsize
\begin{tabular}{c}
\mbox{all paths} \\
\mbox{from $s$ to $x$}\\
\end{tabular}
}}} \;\; \prod_
{\scriptsize
\begin{tabular}{c}
\mbox{all parts (edges)} \\
\mbox{of a path}\\
\end{tabular}
}
A_{\scriptsize \mbox{a part of a path}}
\ee
where $A_{\scriptsize \mbox{a part of a path}}$ denotes the probability amplitude of transition through a part of a given path.

To take into account indistinguishability of different ways of system creation, described by the graph $\Gamma$, we adopting quantum interference (\ref{nslitExper}) to our situation. 
We calculate probability amplitude, as the sum over all different paths of the graph, from minimal to maximal vertex, of product of all edges of the one path, of appropriate fundamental tensor operators.
More precisely, it can be written in terms of formula as\\

\begin{equation}\label{wsp}
\langle f | \lambda t y \rangle =
\hspace{-20pt}
\sum_
{\tiny
\begin{tabular}{c}
\mbox{all different} \\
\mbox{paths from minimal}\\
\mbox{to maximal vertex}\\
\mbox{of the graph}			
\end{tabular}
}
\prod_{
\tiny
\begin{tabular}{c}
\mbox{all edges} \\
\mbox{of the one path}\\
\mbox{of the graph}			
\end{tabular}
}
\left.
 \begin{picture}(1,1)
  \raisebox{2pt}{ \put(0,0){\line(1,3){13}}  \put(0,0){\line(1,-3){13}} }
\end{picture}
\;
{\small
\begin{array}{c}
  \mbox{[$m$]}_n + e_n(\tau_n) \\
  \mbox{[$m$]}_{n-1} + e_{n-1}(\tau_{n-1}) \\
  \vdots \\
   \mbox{[$m$]}_k + e_k(\tau_k)\\
  (m)_{k-1}\\
   \end{array}
   }
 \right |
 \hat F_{k, \tau_n}
\left |
{\small
\begin{array}{l@{}}
  [m]_n  \\
  \mbox{[$m$]}_{n-1} \\
 \vdots \\
   \mbox{[$m$]}_k \\
  (m)_{k-1}\\
   \end{array}
   }
 \right. \;\;\,
 \begin{picture}(1,1)
  \raisebox{2pt}{ \put(0,0){\line(-1,3){13}}  \put(0,0){\line(-1,-3){13}} }
\end{picture}
\end{equation}
 
where $\hat F_{k, \tau_n}$  is the fundamental tensor operator (\ref{fso}), $k=f(j)$ and 
$\tau_n=row(\lambda_{1..j}\setminus \lambda_{1..j-1})$.
Equation (\ref{wsp}) allows to calculate the probability amplitudes of Schur-Weyl state $| \lambda t y \rangle$ in the magnetic configurations $f$ representation.

~~\\~~
\noindent
\textbf{Example}

Let us consider representation system consisting of $N=4$ nodes and single node spin $s=1$ ($n=3$), prepared in the Schur-Weyl state:
\begin{equation}\label{stateExample}
\begin{array}{l}
\Big | \;  \lambda =(3,1), t~=~{\scriptsize \Yvcentermath1 \young(123,2)}, y~=~{\scriptsize \Yvcentermath1 \young(134,2)}   \Big\rangle = 
{\scriptsize \mathbf{\frac{\sqrt{3}}{6}} \; |1,3,2,2\rangle} +  
{\scriptsize \frac{\sqrt{3}}{4}\; |1,2,2,3\rangle} + 
\vspace{4pt}
\\ 
{\scriptsize \frac{\sqrt{3}}{4}\; |1,2,3,2\rangle} -
{\scriptsize \frac{\sqrt{3}}{4}\; |2,1,2,3\rangle} -
{\scriptsize \frac{\sqrt{3}}{6}\; |3,1,2,2\rangle} -
{\scriptsize \frac{\sqrt{3}}{4}\; |2,1,3,2\rangle} -
\vspace{4pt}
\\ 
{\scriptsize \frac{\sqrt{3}}{12}\; |2,3,1,2\rangle} + 
{\scriptsize \frac{\sqrt{3}}{12}\; |3,2,1,2\rangle} -
{\scriptsize \frac{\sqrt{3}}{12}\; |2,3,2,1\rangle} +     
{\scriptsize \frac{\sqrt{3}}{12}\; |3,2,2,1\rangle}.
\end{array} 
\end{equation}
As an example, we show how to calculate the first probability amplitude (marked as bold) of the state (\ref{stateExample}), i.e.
$$
\Big\langle 1,3,2,2      \; \Big | \;      (3,1), {\scriptsize \Yvcentermath1 \young(123,2)}, {\scriptsize \Yvcentermath1 \young(134,2)}   \Big\rangle.
$$
According to presented algorithm we read the sequence\\ $\lambda_{RS} = (\lambda_{1234}=(3,1), \lambda_{123}=(2,1), \lambda_{12}=(1,1), \lambda_{1}=(1))$ from the reversed RSK algorithm.
Next, we construct the graph $\Gamma$

$$
\begin{array}{ccccc}
&
{\tiny
\left (
\begin{array}{@{}c@{}c@{}c@{}c@{}c@{}}
  0 &   & 0 &   & 0  \\
    & 0 &   & 0 &   \\
    &   & 0 &  &  \\
 \end{array}
 \right )}

 &&&\\
 \vspace{-3pt}
 &&\\
 \vspace{-3pt}
 &\downarrow {\tiny 1}&\\
 &&\\
 &
{\tiny
\left (
\begin{array}{@{}c@{}c@{}c@{}c@{}c@{}}
  1 &   & 0 &   & 0  \\
    & 1 &   & 0 &   \\
    &   & 1 &  &  \\
 \end{array}
 \right )}
 &&&\lambda_1 =(1,0,0) \\
 &&\\
 &\downarrow {\tiny 3}&\\
 &&\\
 &
{\tiny
\left (
\begin{array}{@{}c@{}c@{}c@{}c@{}c@{}}
  1 &   & 1 &   & 0  \\
    & 1 &   & 0 &   \\
    &   & 1 &  &  \\
 \end{array}
 \right )}
 &&&\lambda_{12} =(1,1,0) \\
 &&\\
 \swarrow {\tiny 2}&&\searrow {\tiny 2}\\
 &&\\
{\tiny
\left (
\begin{array}{@{}c@{}c@{}c@{}c@{}c@{}}
  2 &   & 1 &   & 0  \\
    & 2 &   & 0 &   \\
    &   & 1 &  &  \\
 \end{array}
 \right )}
 &&
 {\tiny
 \left (
\begin{array}{@{}c@{}c@{}c@{}c@{}c@{}}
  2 &   & 1 &   & 0  \\
    & 1 &   & 1 &   \\
    &   & 1 &  &  \\
 \end{array}
 \right )}
 &&
 \lambda_{123} =(2,1,0) \\
 &&\\
 \searrow {\tiny 2}&&\swarrow {\tiny 2}\\
 &&\\
 &
{\tiny
\left (
\begin{array}{@{}c@{}c@{}c@{}c@{}c@{}}
  3 &   & 1 &   & 0  \\
    & 2 &   & 1 &   \\
    &   & 1 &  &  \\
 \end{array}
 \right )}
 &&&\;\;\;\;\;\; \lambda_{1234} = \lambda =(3,1,0). \\
\end{array}
$$
~~\\~~\\

For the graph presented above, we can read amplitude as ,,sum over two paths''

$$
\Big \langle {\scriptsize (1,3,2,2)} \Big | {\scriptsize (3,1)}, {\scriptsize \Yvcentermath1 \young(123,2)}, \; {\scriptsize \Yvcentermath1 \young(134,2)} \Big\rangle =
$$
$$
{\tiny
\left <
\begin{array}{@{}c@{}c@{}c@{}c@{}c@{}}
  1 &   & 1 &   & 0  \\
    & 1 &   & 0 &   \\
    &   & 1 &  &  \\
 \end{array}
\right |
t_{32}
\left |
\begin{array}{@{}c@{}c@{}c@{}c@{}c@{}}
  1 &   & 0 &   & 0  \\
    & 1 &   & 0 &   \\
    &   & 1 &  &  \\
 \end{array}
\right >
\;
\left <
\begin{array}{@{}c@{}c@{}c@{}c@{}c@{}}
  2 &   & 1 &   & 0  \\
    & 2 &   & 0 &   \\
    &   & 1 &  &  \\
 \end{array}
\right |
t_{21}
\left |
\begin{array}{@{}c@{}c@{}c@{}c@{}c@{}}
  1 &   & 1 &   & 0  \\
    & 1 &   & 0 &   \\
    &   & 1 &  &  \\
 \end{array}
\right >
\;
\left <
\begin{array}{@{}c@{}c@{}c@{}c@{}c@{}}
  3 &   & 1 &   & 0  \\
    & 2 &   & 1 &   \\
    &   & 1 &  &  \\
 \end{array}
\right |
t_{21}
\left |
\begin{array}{@{}c@{}c@{}c@{}c@{}c@{}}
  2 &   & 1 &   & 0  \\
    & 2 &   & 0 &   \\
    &   & 1 &  &  \\
 \end{array}
\right >
}
\; +
$$
$$
{\tiny
\left <
\begin{array}{@{}c@{}c@{}c@{}c@{}c@{}}
  1 &   & 1 &   & 0  \\
    & 1 &   & 0 &   \\
    &   & 1 &  &  \\
 \end{array}
\right |
t_{32}
\left |
\begin{array}{@{}c@{}c@{}c@{}c@{}c@{}}
  1 &   & 0 &   & 0  \\
    & 1 &   & 0 &   \\
    &   & 1 &  &  \\
 \end{array}
\right >
\;
\left <
\begin{array}{@{}c@{}c@{}c@{}c@{}c@{}}
  2 &   & 1 &   & 0  \\
    & 1 &   & 1 &   \\
    &   & 1 &  &  \\
 \end{array}
\right |
t_{21}
\left |
\begin{array}{@{}c@{}c@{}c@{}c@{}c@{}}
  1 &   & 1 &   & 0  \\
    & 1 &   & 0 &   \\
    &   & 1 &  &  \\
 \end{array}
\right >
\;
\left <
\begin{array}{@{}c@{}c@{}c@{}c@{}c@{}}
  3 &   & 1 &   & 0  \\
    & 2 &   & 1 &   \\
    &   & 1 &  &  \\
 \end{array}
\right |
t_{21}
\left |
\begin{array}{@{}c@{}c@{}c@{}c@{}c@{}}
  2 &   & 1 &   & 0  \\
    & 1 &   & 1 &   \\
    &   & 1 &  &  \\
 \end{array}
\right >
}
\; =
$$
$$
{\tiny
\left(\frac{1}{\sqrt{2}}\right)\left(\frac{1}{\sqrt{2}}\right)\left(\frac{\sqrt{3}}{12}\right) + \left(\frac{1}{\sqrt{2}}\right)\left(\frac{1}{\sqrt{6}}\right)\left(\frac{3}{4}\right)= \; \frac{\sqrt{3}}{6}
}
$$
One can check, that standard method (\ref{spr20}) gives the same result.

\section{Concluding remarks}

We have presented a new method of Schur-Weyl states generation for quantum spin system.
Using this method we can calculate probability amplitudes in a very efficient way using only simple combinatorial operations. 
In contrast to the standard method \cite{bohr_mottelson} it is size independent of the physical system and it is well adopted to the computer implementation. 

This method is devoted to the researchers who investigate symmetry of quantum systems consisting of many identical subsystems and want to reduce the size of eigenproblem, or in general, diminish the representation matrix of any physical quantities, represented in the symmetric or unitary group algebra.
Clearly, the algorithm can be used for appropriate subsets of irreducible bases, in particular for an inhomogeneous models, with the weight $\mu$, being a partition which differs from a rectangular shape.
Another property of the Schur-Weyl states is the fact, that representation of physical systems in Hilbert space spanned on such states enable to extract information hidden in nonlocal degrees of freedom. This feature can be very useful in broad range of problems in Quantum Computations, especially in quantum algorithms constructions.

The novelty of the proposed algorithm and also the main idea behind it, is the fact, that while constructing the quantum state via addition of consecutive nodes, we are dealing with a ,,combinatorial quantum interference'' of all the possible ways of addition the new node to an already existing system. This observation leads to formula (\ref{wsp}) which significantly simplifies the calculation of probability amplitudes of Schur-Weyl states.
All operations which constitute the algorithm are reduced to simple arithmetic operations such as  multiplication and summation and hence, the algorithm is polynomial in time with respect to  $N$ and $n$, in contrary to the standard method \cite{bohr_mottelson} which uses the summation over the symmetric group, and thus grows exponentially with $N$.

\providecommand{\href}[2]{#2}

\end{document}